\documentclass[aps,preprint,preprintnumbers,nofootinbib,showpacs]{revtex4}

\usepackage{amsmath,amsfonts,amssymb}
\usepackage{color,graphicx}
\usepackage{cancel}
\usepackage{appendix}
\usepackage{mathtools}
\usepackage{dsfont}

\def\om{\omega}

\def\bea{\begin{eqnarray}}
\def\eea{\end{eqnarray}}
\def\sea{\nonumber \\ &&}

\def\zb{\beta}
\def\zc{\gamma}

\def\ssc{\scriptscriptstyle}
\def\lsim{\mathrel{\raise.3ex\hbox{$<$\kern-.75em\lower1ex\hbox{$\sim$}}} }
\def\gsim{\mathrel{\raise.3ex\hbox{$>$\kern-.75em\lower1ex\hbox{$\sim$}}} }

\newcommand{\verteq}{\rotatebox{90}{$\,=$}}
\newcommand{\equalto}[2]{\underset{\scriptstyle\overset{\mkern4mu\verteq}{#2}}{#1}}

\begin{document}


\vspace*{0.5in}
\title{Special Relativity and its Newtonian Limit from a Group Theoretical Perspective}


\author{\bf Otto C. W. Kong  and Jason Payne
}
\email{otto@phy.ncu.edu.tw}

\affiliation { Department of Physics and  Center for High Energy and High Field Physics, 
National Central University, Chung-li, Taiwan 32054  \\
}

\begin{abstract}\vspace*{.2in}
In this pedagogical article, we explore a powerful language for 
describing the notion of spacetime and particle dynamics in it 
intrinsic to a given fundamental physical theory, focusing on  
special relativity and its Newtonian limit. The starting point 
of the formulation is the representations of the relativity 
symmetries. Moreover, that seriously furnishes -- via the notion 
of symmetry contractions --  a natural way in which one can 
understand how the Newtonian theory  arise as an approximation 
to Einstein's theory. We begin with the Poincar\'{e} symmetry
underlying special relativity and the nature of Minkowski
spacetime as a coset representation space of the algebra and 
the group. Then, we proceed to the parallel for the phase space
of a  spin zero particle, in relation to which we present the full 
scheme for its dynamics under the Hamiltonian formulation 
illustrating that as essentially the symmetry feature of the phase 
space geometry. Lastly, the reduction of all that to the Newtonian 
theory as an approximation with its space-time, phase space, and 
dynamics under the appropriate relativity symmetry contraction 
is presented. While all notions involved are well established,
the systematic presentation of that story as one coherent
picture fills a gap in the literature on the subject matter.
\end{abstract}

\maketitle

\section{Introduction}
Over the past century, the notion of symmetry has become an 
indispensable feature of theoretical physics. It no longer merely 
facilitates the simplification of a difficult calculation, nor 
lurks behind the towering conservation laws of Newton's time, but 
rather unveils fundamental features of the universe around us, 
describes how its basic constituents interact, and places deep 
constraints on the kinds of theories that are even possible. In 
light of this, one natural approach to contemplating nature 
would be to take symmetries {seriously}. In other words, 
one may aim to reformulate as much of our understanding of 
nature in the language most natural for describing symmetries.

More specifically, what we will be concerned with here is the 
notion of a \textit{relativity symmetry}. Using the archetypal 
example handed down to us by Einstein, we hope to illustrate 
that not only does a relativity symmetry relate frames of 
reference in which the laws of physics look the same, but 
also captures the structure of physical spacetime\footnote{
Throughout this note we will reserve ``spacetime'' specifically 
for the notion of physical spacetime underlying Einsteinian 
special relativity, while using ``space'' to cover the 
corresponding notion in general. Spacetime in the Newtonian 
setting, or space-time used instead whenever admissible, 
refers to the sum of the mathematically independent 
Newtonian space and time.
} 
itself as well as much of the theory of particle dynamics on 
it. Moreover, as we will detail below, formulating our theory 
in these terms provides one with a natural language in which 
approximations to the theory in various limits can be 
described. The term relativity symmetry, 
though much like introduced into physics by Einstein, is in 
fact a valid notion for Newtonian mechanics too. It just has
a different relativity symmetry, the Galilean symmetry.
Newtonian mechanics can be called the theory of  Galilean 
relativity. The symmetry for the Einstein theory is usually
taken as the Poincar\'e symmetry. Note that we neglect the 
consideration of all discrete symmetries like the parity
transform in this article.  We talk about the relativity 
symmetries without such symmetries included. 

In order to parse the details of this fascinating tale, we 
begin with an examination of exactly how one can naturally 
pass from the (classical) relativity symmetry group/algebra 
to its corresponding geometric counterparts such as the 
model of the spacetime and the phase space for a particle 
in Section II, for the case of the Poincar\'e symmetry $ISO(1,3)$. 
The full dynamical theory, for spin zero case, under the Hamiltonian
formulation is to be presented from the symmetry perspective 
in Section III. From there, in Section IV, we provide a brief 
introduction to the language of approximating a symmetry: 
contractions of Lie algebras/groups, and their representations. 
This is augmented by a continuation of the exploration of 
special relativity, and in particular the way in which the 
Newtonian limit is to be understood within this context,
before giving some concluding remarks in the last section.  
Finally, we have added an appendix  (Appendix~A) discussing 
what is, in essence, the reverse procedure of what is described 
in this paper: symmetry deformations. This is another 
fascinating facet of the overall story, which provides one with 
a reasonable procedure for determining what sort of theories 
a given theory may be an approximation of. We could have 
discovered Einstein's Theory of Special Relativity from 
the deformation of Galilean relativity if we had the idea
 \cite{D,Min}. Historically, however, 
careful studies of symmetries and their role in physics 
barely started in the time of Einstein, and only began to 
pick up momentum with the development of quantum 
mechanics in the 1930s.

The basic material treated here is, in our opinion, important 
for a good appreciation of the physical theories, yet perhaps 
not as well-known as it should be. The key parts of the 
presentation have, apparently, not been otherwise explicitly 
available in the literature, not to say the full story in one place. 
Hence, we make this effort to present it, aiming at making it 
accessible even to students with limited background. For the 
latter purpose, we include an extra appendix ({Appendix}~B)
giving a physicist's sketch of necessary group theory background, 
to make the article more self-contained. We see the presentation 
as useful to physicists in a couple of ways. Firstly, for the existing 
theories, it gives a coherent and systematic way to organize all 
aspects of the theories within one framework, highlighting their 
mutual relationship. That can improve our understanding of all 
aspects of the theoretical structure. Second, a particular way to 
look at a theory, even if not in any sense superior to the other 
ways, may provide a specific channel to go to theories beyond. 
The authors' attention on the subject matter
is closely connected to our recent studies essentially on the
exact parallel constructions for the theories of quantum mechanics, 
including retrieving of the `nonrelativistic' from the `relativistic' 
as well as classical from quantum  \cite{070,087}. The Lorentz 
covariant theory of quantum mechanics resulted is new, with
a quantum notion of Minkowski metric. A better understanding 
of that actually gives also a notion of Newtonian mass and a new 
insight into the Einstein on-shell mass condition, to be reported
in a forthcoming article \cite{094}.  The symmetry for the  
`nonrelativistic' quantum mechanics is essentially a $U(1)$ central 
extension \cite{Gil,gq} of the Galilean symmetry, or the one with the 
Newtonian time translation taken out -- we called that $H_{\!\ssc R}(3)$
as the Heisenberg-Weyl symmetry with three noncommuting
$X$-$P$ pairs supplemented by the $SO(3)$ \cite{070,094}. 
For the `relativistic' case, we found it necessary to go beyond
the Poincar\'e symmetry to the larger $H_{\!\ssc R}(1,3)$
  \cite{087,094}. The last reference also addresses nonzero spin 
and composite systems. Details of those are beyond the scope 
of the present article. All that illustrate well the value of looking 
at the well known theories from a somewhat different point of 
view seriously, as done here.

\section{From Relativity to Physical Spacetime and the Particle Phase Space}
A conventional path to the formulation of a physical 
theory is to start with a certain collection of assumptions 
about the geometry of the physical spacetime objects in this 
theory occupy. That is to say, the theory starts with taking 
a mathematical model for the intuitive notions of the 
physical space and time. After all, dynamics means the study  
of motion, which is basically the change of position with
respect to time. In Newtonian mechanics, Newton himself 
followed the basic definitions in his \textit{Principia} with 
his Scholium arguing for Euclidean space coupled with 
absolute time as the foundation of the description of the 
physical world; the study of special relativity may be 
introduced via Minkowski spacetime; general relativity
typically assumes the universe is a (torsion-free) Lorentzian 
manifold, and the list goes on. It is then \textit{from} this 
`foundation' that one infers the symmetries present in the 
model. Note that in Newton's time, Euclidean geometry 
is really the only geometry known. What we hope to 
convince the reader of in this section is that the opposite 
path can be just as fruitful, if not more so. In particular, 
we will \textit{start} with the relevant (relativity) 
symmetry, given by a Lie group (and its associated Lie 
algebra), and couple it to the representation that naturally 
captures the underlying geometry. Once the basic
definitions are in place, we use special relativity as an
illustrative example of this procedure. The approach 
will be extended to present the full theory of particle 
dynamics in the next section. Note that the model for 
the physical space or spacetime is closely connected to 
the theory of particle dynamics on it. First of all, Newton
introduced the notion of particle as point-mass to serve
as the ideal physical object which has a completely
unambiguous position in his model of the physical space.
Conversely, in a theory of
particle dynamics, there is no other physical notion of the 
physical space itself rather than the collection of possible 
positions for a free particle (or the center of mass for 
a closed system of particles which, however, have to 
be defined based on the full particle theory, for example
the three Newton's Laws). It is and has to be the 
configuration space for the free particle. 

\subsection{The Coset Space Representation}

In his seminal paper \textit{Raum und Zeit}
 \cite{Min}, Hermann Minkowski famously said, 
\begin{quote}The views of space and time which 
I wish to lay before you have sprung from the soil 
of experimental physics, and therein lies their 
strength. They are radical. Henceforth space by 
itself, and time by itself, are doomed to fade away 
into mere shadows, and only a kind of union of 
the two will preserve an independent reality.
\end{quote}
 In this statement, Minkowski reveals 
something of tremendous importance: the idea of
Lorentz symmetry as the right  transformations sending 
inertial frames to inertial frames directly alter the model
geometry of the physical space and time, or spacetime,
from the Newtonian theory. The model for the physical 
spacetime itself depends on the explicit form of the
 \textit{Principle of Relativity} being postulated,
{\em i.e.} the relativity symmetry of the theory. In this 
subsection, we will take this realization to heart and 
explore precisely how one goes about recovering the 
model for the physical spacetime naturally associated 
with a given relativity structure for the classical theories.

Consider a Lie group $G$, with associated Lie 
algebra $\mathfrak{g}$, which we take as 
capturing the finite and infinitesimal transformations, 
respectively, that we can perform on a given physical 
system without changing the form of the physical 
laws. In other words, those transformations which 
take a given (inertial) frame of reference into another 
equally valid frame. $G$ is then the relativity 
symmetry, or the symmetry group of the spacetime 
model of the theory of particle dynamics.

The use of the word ``transformation'' above already 
hints at the need for a representation-theoretic 
perspective of what, exactly, the relativity symmetry 
encodes. Indeed, as it stands the mathematical group
$G$ is merely an abstract collection of symbols obeying 
certain rules -- a representation capturing the group 
structure is required to illuminate what these rules 
really mean in terms of \textit{physical transformations},
which are mathematically transformations on a vector
space. The best examples of the latter are our Minkowski
spacetime and the Newtonian space-time. The first,
perhaps prosaic, step in this direction is simply to use 
the group multiplication, thought of as a (left) action 
of $G$ on itself:
\[g'\cdot g \mapsto g'g.\]
In other words, we can try to imagine that what we 
mean by a location/position in the ``physical spacetime'' 
is nothing more than an element $g\in G$, and that a 
transformation is then simply furnished directly by the 
group operation. We have at hand the Poincar\'e symmetry
denoted by $ISO(1,3)$ consisting of the rotations and
translations conventional defined as isometries
of the Minkowski spacetime. However, to conform completely
to the perspective of taking the symmetry group as the starting 
point, we are going to simply see the group as the Lie group 
obtained from the corresponding Lie algebra $\mathfrak{iso}(1,3)$ 
presented in Eq.(\ref{PS}) below as abstract mathematical objects.  
We can take each element of the pure 
translations as a point in the Minkowski spacetime,
which is equivalent to saying that each point is to be
identified as where you get to after a particular spacetime
translation from the origin. Note that while the rotations
take any point other than the origin to a different point,
they do not move the origin. From the abstract 
mathematical point of view, what we described here is
called a coset space. The Minkowski spacetime is a coset
space of the Poincar\'e symmetry. 

From here we consider the coset space $M:=G/H$,
defined  mathematically as like a quotient of the group
$G$ by a closed subgroup $H<G$. A coset $gH$ containing
the element $g$ is the collection of all group elements
of the form $gh$ where $h$ is any element in $H$. Note
\[
g'H = g(g^{-1}g' H) =gH \qquad \mbox{for} \quad  g^{-1}g' \in H \;.
\]
Observe that the above action descends to an action 
of the full group $G$ on $M$ in an obvious way as
\[g'\cdot (gH)=(g'g)H \;.\]
It is more convenient to use the Lie algebra notation.
We write a group element in terms of 
\[g=\exp( a^i X_i)\;,\]
where the $X_i$ are the generators and $a^i$ real
parameters (note that, as is typical, we are using 
the Einstein summation convention). $X=a^i X_i$ as a
linear combination of the generators, as basis elements, 
is an element of the Lie algebra $\mathfrak{g}$.
Each coset then can be conveniently identified with
an element 
\[  \exp(s^j Y_j)  \]
where $Y_j$ are the generators among the $X_i$
set which serves as a basis for the vector subspace
$\mathfrak{p}$ of $\mathfrak{g}$ complementary 
to the subalgebra $\mathfrak{h}$   for $H$, {\em i.e.}
$\mathfrak{g}=\mathfrak{h} +\mathfrak{p}$
as a vector space. The real numbers $s^j$ can be
seen as coordinates for each coset as a point in the 
coset space (space of the cosets) and the group action 
as symmetry transformations on the coset space, or 
equivalently the reference frame transformations. Let 
us look at such a transformation at the infinitesimal 
limit. 

We are going to need a specific form of the
{Baker-Campbell-Hausdorff} (BCH) series for the case of 
products between a coset representative $\exp(Y)$ and 
an infinitesimal element $\exp(\bar{X})$. In particular,
the result
\bea \label{bch2}
\exp(\bar{X})\exp(Y) 
= \exp\!\left( Y - [Y,\bar{X}] \right) \exp(\bar{X})\;,
\eea 
can be easily checked to hold in general, though no 
similarly simple expression can be find for two 
operators/matrices neither infinitesimal, with generic
commutation relation.

\subsection{From the Poincar\'{e} Algebra to Minkowski Space}
The protagonists of our story are the Poincar\'{e} group 
and algebra $ISO(1,3)$ and $\mathfrak{iso}(1,3)$. 
These describe the finite and infinitesimal 
transformations, respectively, that turn one (relativistic) 
inertial frame into another, i.e. the symmetry which puts 
the ``relativity'' in Einstein's special relativity. Recall 
that the Lie algebra $\mathfrak{iso}(1,3)$ possesses ten 
generators, which are split up into the six generators of 
rotations, among the spacetime directions, $J_{\mu\nu}$ 
(where $0\leq\mu<\nu\leq 3$) and the four generators 
of translations along the four directions\footnote{
The conventional description of $\mathfrak{iso}(1,3)$ 
uses instead the ``momentum"  $P_\mu$ as generators, 
which are related to the generators as``energy" used here 
by $E_\mu=cP_\mu$. As we will see in the following 
sections, $E_\mu$ are the more natural choice from the 
perspective of symmetry contractions.}
$E_\mu$, and which satisfy the following commutation 
relations\footnote{
In the mathematicians' notation, the commutator is
really the Lie product defining the real Lie algebra to
which the set of generators is a basis more naturally
without all the $i\hbar$. Physicists version among to 
rescaling all the generators by the $i\hbar$ factor,
 the mathematically unreasonable $i$ to have the
generators correspond (in a unitary representation) 
to physical observables and $\hbar$ to give the proper 
(SI) units to them. Strictly speaking, we should be
thinking about  like
$-\frac{i}{\hbar} E_\mu $ and $-\frac{i}{\hbar} J_{\mu\nu}$
as our basis vectors are, {\em i.e.} the true generators, of the 
real Lie algebra, which is the real linear combination of them,
with parameters in the proper physical dimensions.}:
\begin{align}
&[J_{\mu\nu},J_{\lambda\rho}]=
- i\hbar(\eta_{\nu\lambda}J_{\mu\rho}-\eta_{\mu\lambda}J_{\nu\rho}
+\eta_{\mu\rho}J_{\nu\lambda}-\eta_{\nu\rho}J_{\mu\lambda})\;, 
\nonumber \\
&
[J_{\mu\nu},E_{\rho}]=
- i\hbar(\eta_{\nu\rho}E_{\mu}-\eta_{\mu\rho}E_{\nu})\;,
\qquad [E_{\mu},E_{\nu}]=0 \;,
\label{PS}
\end{align}
 with $J_{\mu\nu}$ with $\mu>\nu$ to be interpreted 
as  $-J_{\nu\mu}$, and we use $\eta_{\mu\nu}=\{-1,1,1,1\}$
as like the Minkowski metric. For easy reference, we take 
a notation convention which is essentially the same 
as that of the popular text book by Tung \cite{T}, besides 
using $E_\mu$ and an explicit $\hbar$.

It is intuitively clear (and easy to check) that the subset
$\mathfrak{so}(1,3)$ generated by the $J_{\mu\nu}$ 
generators forms a subalgebra of $\mathfrak{iso}(1,3)$
 -- the subalgebra of spacetime rotations called Lorentz
transformations. Thus, if we are interested in the coset 
representation introduced in the previous section, the 
candidate for our Minkowski spacetime should be the 
coset space $\mathfrak{M}:=ISO(1,3)/SO(1,3)$. 
We write a generic element $X\in\mathfrak{iso}(1,3)$ 
and $Y\in\mathfrak{iso}(1,3)- \mathfrak{so}(1,3)$ 
(as the complementary space $\mathfrak{p}$) as
\[
X= -\frac{i}{\hbar} \left( \frac{1}{2}\omega^{\mu\nu}J_{\mu\nu}+b^{\mu}E_{\mu} \right) 
\qquad\textnormal{and}
\qquad Y= -\frac{i}{\hbar}t^{\rho}E_{\rho} \;,
\]
respectively. Note that we have put in a factor of
$\frac{1}{2}$ in the sum $\om^{\mu\nu}J_{\mu\nu}$,
with $\om^{\mu\nu}=-\om^{\nu\mu}$, to lift the
$\mu<\nu$ condition for convenience. Distinct 
elements in the form $Y$ are in one-to-one
correspondence with the distinct cosets.
Next, as we saw in the preceding discussion, we will 
pass from this to an action on the corresponding coset 
space $\mathfrak{M}$ (which, as we will see below, 
is isomorphic to Minkowski space, $\mathds{R}^{1,3}$). 
Consider an infinitesimal transformation given in the 
group notation as $g'=\exp(\bar{X}_{\!\ssc H} + \bar{Y})
  = 1 +\bar{X}_{\!\ssc H} + \bar{Y} =\exp(\bar{Y})\exp(\bar{X}_{\!\ssc H})$,  
with $\bar{X}_{\!\ssc H} = -\frac{i}{2\hbar}\bar\om^{\mu\nu}J_{\mu\nu}$
and $\bar{Y} = -\frac{i}{\hbar}\bar{t}^{\mu} E_{\mu}$. 
We first check that
\begin{align*}
[\bar{X}_{\!\ssc H},Y] 
&= -\frac{1}{2\hbar^2}\bar\omega^{\mu\nu}t^{\rho}[J_{\mu\nu},E_{\rho}]\\
&= \frac{i}{2\hbar} t^{\rho}(\bar\omega^{\mu\nu} \eta_{\nu\rho}E_{\mu}+\bar\omega^{\nu\mu}\eta_{\mu\rho}E_{\nu}) \\
&= \frac{i}{\hbar} \bar\omega^{\mu}_{\,\rho}t^{\rho}E_{\mu}\;,
\end{align*}
 and $[Y, [\bar{X}_{\!\ssc H},Y]] =0$.  Applying our BCH formula 
(\ref{bch2}) for the case, we have
\begin{align*}
\exp(\bar{X}_{\!\ssc H})\exp(Y) 
&= \exp(Y-[Y,\bar{X}_{\!\ssc H}]) \exp(\bar{X}_{\!\ssc H}) \\
&=  \exp( [\bar{X}_{\!\ssc H},Y]) \exp(Y) \exp(\bar{X}_{\!\ssc H})
\end{align*}
as exact in the infinitesimal parameters in $\bar{X}_{\!\ssc H}$.
Thus, the multiplication $g'\cdot(gSO(1,3))$ yields 
\begin{align*}
& \exp\!\left( -\frac{i}{\hbar}\bar{t}^{\mu} E_{\mu} \!\right) 
    \exp\!\left( -\frac{i}{2\hbar}\bar\om^{\mu\nu}J_{\mu\nu} \!\right) 
    \exp\!\left( -\frac{i}{\hbar}t^{\rho}E_{\rho} \!\right) SO(1,3) \\
& =\exp\!\left( -\frac{i}{\hbar} \textcolor{red}{ \bar{t}^{\mu} E_{\mu} } \!\right) 
   \exp\!\left(  \frac{i}{\hbar} \textcolor{red}{ \bar\omega^{\mu}_{\,\rho}t^{\rho}E_{\mu} } \!\right) 
    \exp\!\left( -\frac{i}{\hbar} \textcolor{blue}{t^{\rho}E_{\rho} }\!\right) 
  \textcolor{green}{ \exp\!\left( -\frac{i}{2\hbar}\bar\om^{\mu\nu}J_{\mu\nu} \!\right) } SO(1,3) \\
&= \exp\!\left( \!-\frac{i}{\hbar}\bigg(\textcolor{blue}
  {\equalto{t^{\mu}E_{\mu}}{\textnormal{original } t^{\mu} \textnormal{ part}}}
  +\textcolor{red}{\equalto{( \bar{t}^{\mu} -\bar\omega^{\mu}_{\,\rho}t^{\rho}
    ) E_{\mu}}{\textnormal{infinitesimal change}}} \bigg)\!\!\right) 
    \textcolor{green}{SO(1,3)} \;,
\end{align*}
which is the resulted coset of 
\[
 \exp\!\left( -\frac{i}{\hbar} (t^{\mu} +dt^{\mu}) E_{\mu} \!\right) {SO(1,3)}
\]
where the infinitesimal change in coordinate $t^{\mu}$
is given by  
$dt^{\mu}= -\bar\omega^{\mu}_{\;\nu}t^{\nu}  +\bar{t}^{\mu}$.
The last equation can be seen as giving a representation 
of $\mathfrak{iso}(1,3)$ on $\mathfrak{M}$ by 
identifying the coset represented by $Y$ with the column vector 
$(t^{\mu},1)^{\ssc T}$ and $\bar{X}= \bar{X}_{\!\ssc H}+\bar{Y}$ 
with the matrix:
\[
\bar{X}= \frac{i}{\hbar}\big(-\bar\omega^{\mu\nu} J_{\mu\nu}
  +\bar{t}^{\mu}  E_{\mu}\big) \xrightarrow{\textnormal{represented by}}
   \left( \begin{array}{cc}
-\bar\omega^{\mu}_{\;\nu} & \bar{t}^{\mu}  \\
0 & 0 \\
\end{array}  \right)
\]
so that 
\bea\label{t-trans}
\left( \begin{array}{c}
dt^{\mu} \\ 
0 \\
\end{array}  \right)
= \left( \begin{array}{cc}
-\bar\omega^{\mu}_{\;\nu} & \bar{t}^{\mu}  \\
0 & 0 \\
\end{array}  \right)
\left( \begin{array}{c}
t^{\nu} \\ 
1 \\
\end{array}  \right)
=\left( \begin{array}{c}
 -\bar\omega^{\mu}_{\;\nu}t^{\nu}  +\bar{t}^{\mu} \\
0 \\
\end{array}  \right)\;.
\eea 

We have derived above the representation of the Lie algebra 
$\mathfrak{iso}(1,3)$ for the infinitesimal transformations 
of the coset space $\mathfrak{M}$ which obviously can be 
seen as a vector space with $t^\mu$ being the four-vector. 
The elements of $\mathfrak{iso}(1,3)$ associated with the 
infinitesimal transformations with $\bar{t}^{\mu}=0$,
 {\em i.e.} elements of the Lorentz subalgebra 
$\mathfrak{so}(1,3)$, indeed exponentiate into a $SO(1,3)$ 
Lorentz transformation on $t^\mu$ as
\begin{align*}
\left( \begin{array}{cc}
-\omega^{\mu}_{\;\nu} & 0 \\ 
0 & 0
\end{array}  \right) 
&\xrightarrow{\quad\exp\quad} 
\left( \begin{array}{cc}
\Lambda^{\mu}_{\;\nu} & 0 \\ 
0 & 1
\end{array}  \right) \\
&\xrightarrow[\textnormal{the action}]{\textnormal{leads to}}
\left( \begin{array}{cc}
\Lambda^{\mu}_{\;\nu}  & 0 \\ 
0 & 1
\end{array}  \right)
\left( \begin{array}{c}
t^{\nu} \\ 
1
\end{array}  \right) 
 =\left( \begin{array}{c}
\Lambda^{\mu}_{\;\nu} t^{\nu}\\ 
1
\end{array}  \right)\;.
\end{align*}
Similarly, the infinitesimal translations
exponentiate into the finite translations
\[
\exp \!\left( \begin{array}{cc}
0 & b^{\mu}  \\
0 & 0 \\
\end{array}  \right)
=  \left( \begin{array}{cc}
\delta^{\mu}_{\nu} & B^{\mu} \\ 
0 & 1
\end{array}  \right) \;.
\]
In fact, the Poincar\'e symmetry is given in 
physics textbooks typically as the transformations
\[
x^\mu  \rightarrow \Lambda^{\mu}_{\;\nu} x^\nu  + A^{\mu} \;.
\]
from which one can obtained the same
infinitesimal transformations with
$d (\Lambda^{\mu}_{\;\nu}) = (-\omega^{\mu}_{\;\nu})$
and  $\frac{1}{c} dA^{\mu}$ similarly associated with $b^{\mu}$,
switching from $x^\mu$ to our $t^\mu= \frac{1}{c} x^\mu$.
That is actually defining a symmetry group through 
a representation of its generic element. Putting that
in the matrix form, we have
\begin{align*}
\left( \begin{array}{c}
\Lambda^{\mu}_{\;\nu} t^{\nu} + B^\mu\\ 
1
\end{array}  \right)
&=  \left[ \left( \begin{array}{cc}
\delta^{\mu}_{\rho} & B^{\mu} \\ 
0 & 1
\end{array}  \right)
\left( \begin{array}{cc}
\Lambda^{\rho}_{\;\nu}  & 0 \\ 
0 & 1
\end{array}  \right) \right]
\left( \begin{array}{c}
t^{\nu} \\ 
1
\end{array}  \right)  \\
&= \exp \!\left( \begin{array}{cc}
0 & b^{\mu}  \\
0 & 0 \\
\end{array}  \right)
\exp \!
\left( \begin{array}{cc}
-\omega^{\rho}_{\;\nu} & 0 \\ 
0 & 0
\end{array}  \right) 
\left( \begin{array}{c}
t^{\nu} \\ 
1
\end{array}  \right)  \;,
\end{align*}
from which we can see the infinitesimal limit
of the transformation matrix being
\[
\left[ I + \left( \begin{array}{cc}
0 & b^{\mu}  \\
0 & 0 \\
\end{array}  \right) \right]
\left[ I + \left( \begin{array}{cc}
-\omega^{\mu}_{\;\nu} & 0 \\ 
0 & 0
\end{array}  \right) \right]
= I + \left( \begin{array}{cc}
-\omega^{\mu}_{\;\nu} &  b^{\mu} \\ 
0 & 0
\end{array}  \right) \;.
\]
In fact, we can think of each point $(t^\mu,1)^{\ssc T}$ 
in $\mathfrak{M}$ as being defined by the action of 
the above matrices on the coordinate origin $(0,1)^{\ssc T}$ 
by taking $B^\mu=t^\mu$. Indeed
\bea
\left(\begin{array}{c}
t^{\mu} \\ 1
\end{array}\right)
\equiv
\left(\begin{array}{cc}
\Lambda^{\mu}_{\,\nu} & t^{\mu} \\
0 & 1
\end{array}\right)
\left(\begin{array}{c}
0 \\ 1
\end{array}\right)
= \left(\begin{array}{c}
  t^\mu\\ 1
\end{array}\right) \; ;
\eea
hence the $t^{\mu}$-space is essentially isomorphic 
to the collection of matrices of the form
\[
\left(\begin{array}{cc}
\Lambda^{\mu}_{\,\nu} & t^{\mu} \\
0 & 1
\end{array}\right)\;.
\]
Then, each of the translational elements can be taken 
as the standard representative for the coset
\[  
\left(\begin{array}{cc}
\delta^\mu_\nu & t^{\mu} \\
0 & 1
\end{array}\right) SO(1,3) \;.
\]
The latter, therefore, describes a full coset and the
vector space of all such cosets is isomorphic to that
of the collection of all 
$e^{\left(-\frac{i}{\hbar} t^\mu E_\mu\right)} SO(1,3)$
from the abstract mathematical description we start with.

When the Minkowski spacetime is taken as the
starting point, it is a homogeneous space in the
physical sense that every point in it is really much the
same as another. Each can be taken as the origin on
which we can put in a coordinate system fixing a
frame of reference. The symmetry of it as a geometric 
space  is caught in the mathematical definition of
a homogeneous space as a space with a transitive
group of symmetry, meaning every two points in it 
can be connected through the action of a group 
element. For a particular point like the origin, there
is a subgroup of the symmetry that does not move
it, which is called the little group. It is a mathematical
theorem that the homogeneous space is isomorphic
to the coset space of the symmetry group ``divided
by" the little group. Our result of the Minkowski 
spacetime as $ISO(1,3)/SO(1,3)$, whether in terms 
of the $t^\mu$ or the $x^\mu$ coordinates, is just 
a case example.

Indeed, using $t^\mu$ as the coset space coordinates
 is really no different from using $P_\mu$ as generators 
and $x^\mu$. This is because we can write Lorentz 
transformations as
\bea
x^{\prime\ssc 0} &=& \gamma ( x^{\ssc 0} + \beta_i x^i) \;
\nonumber \\
x^{\prime i} &=& \gamma ( x^{i} + \beta^i x^{\ssc 0}) \;,
\eea
or equivalently as
\bea
t^{\prime\ssc 0} &=& \gamma ( t^{\ssc 0} + \beta_i t^i) \;
\nonumber \\
t^{\prime i} &=& \gamma ( t^{i} + \beta^i t^{\ssc 0}) \;,
\eea
with $\beta_i=\tfrac{v_i}{c}$, $\beta^i=\tfrac{v^i}{c}$, 
and $\gamma=\frac{1}{\sqrt{1-\beta_i\beta^i}}$.
Both of the above are equivalent to
\bea
t^{\prime} &=& \gamma ( t +\frac{\beta_i}{c} x^i) 
= \gamma ( t +\frac{ v_i}{c^2} x^i)\;
\nonumber \\
x^{\prime i} &=& \gamma ( x^{i} +\beta^i ct) = \gamma ( x^{i} + v^i t)\;,
\eea
where $t\equiv t^{\ssc 0}$. In other words, $t^\mu$ 
and $x^\mu$ describe the same spacetime ``position'' 
four-vector, they are simply expressed in time
and space units, respectively. Einsteinian relativity 
says space and time are coordinates of a single 
spacetime, hence they are naturally to be expressed 
in the same units. It does not say that the spatial 
units are preferable, or in some sense more natural, 
than the time units! Straight to the spirit of special
relativity, we should rather use the same unit to
measure $t^\mu$ and $x^\mu$ in which $c=1$. With 
the different units, although textbooks typically 
use $x^\mu$, what we show below is that we should 
indeed start with $t^\mu$ as coordinates for 
Minkowski spacetime, as we have done above, if we 
want to directly and naturally recover $t$ and $x^i$ 
as coordinates of the representation space of 
Newtonian physics in the Newtonian limit, {\em i.e.} 
under the symmetry contraction described in the
following section. 

In physical terms, $J_{\mu\nu}$ has the units of $\hbar$, 
while the algebra element $-\frac{i}{\hbar}(\omega^{\mu\nu} J_{\mu\nu}+ b^\mu E_\mu)$
has no units (for we do not want to exponentiate something 
that has units). Hence, $\omega^{\mu\nu}$ must also have no 
units, and $b^\mu E_\mu$ has the units of $\hbar$, giving 
$b^\mu$ the unit of time. Similarly, $a^\mu$, and $x^\mu$, 
as well as $A^\mu$, have the units of $\hbar$ divided by 
that of $P_\mu$. All quantities now have the right units, 
and $c$ of course has the units of $\frac{x^\mu}{t^\mu}$,
 {\em i.e.}  distance over time.

\subsection{The Phase Space for Particle Dynamics as a Coset Space}
After the Minkowski spacetime $\mathfrak{M}$ described 
above, we come to another important coset space of the 
Poincar\'e symmetry, one that serves as the phase space 
for a single particle. Besides the spacetime coordinates,
we also need  the momentum or equivalently the velocity
coordinates. However, the only parameters in the
description of the group elements that correspond to
velocity are those for the components of the three-vector 
$\zb^i=\om^{i\ssc 0}$.
The candidate coset space is $ISO(1,3)/SO(3)$ which 
is seven-dimensional. An otherwise candidate is 
$ISO(1,3)/T_{\!\ssc H} \times SO(3)$ where $T_{\!\ssc H}$ 
denotes the one-parameter group of (`time') translations 
generated by $H=E_{0}$, which corresponds to the 
physical energy. That space loses the time coordinate 
$t^{\ssc 0}$ which cannot be desirable. There is a 
further option of extending $ISO(1,3)/SO(3)$. Let us 
first look carefully at the latter coset space. Instead 
of deriving that coset space `representation' from the 
first principle as for the Minkowski spacetime above,
however, we construct it differently. The coset space 
here is not a vector space, hence the group action on 
it is not a representation. Without the linear structure, 
the group transformations cannot be written in terms 
of matrices acting on vectors representing the states 
each as a point in the space. Moreover, obtaining the 
resultant coset of a generic group transformation on 
a coset following the approach above is a lot more 
nontrivial. A vector space description of a phase 
space as a simple extension of the coset space can
be constructed from physics consideration.

Newtonian mechanics as the nonrelativistic limit to
special relativity has of course a six-dimensional 
vector space as the phase space, each point in which
is described by two three-vectors, the position vector
$x^i$ and the momentum vector $p^i$. The two parts are 
in fact independent coset space representations of the 
corresponding relativity symmetry -- the Galilean 
relativity. Or the full phase space can be taken as
a single coset space. Going to special relativity, the 
three-vectors are to be promoted to Minkowski four-vectors. 
A four-vector is an element in the four dimensional
irreducible representation of the $SO(1,3)$ symmetry,
while a  three-vector belongs to the three dimensional
irreducible representation of the $SO(3)$ group as
a subgroup of $SO(1,3)$. Promoting $x^i$ to $x^\mu$
we get the Minkowski spacetime $\mathfrak{M}$ depicted
with $t^\mu=\frac{x^\mu}{c}$ as the $ISO(1,3)/SO(1,3)$ 
coset space. Things for the momentum four-vector $p^\mu$ 
are somewhat different. It is a constrained vector with 
magnitude square $p_\mu p^\mu$ fixed by the particle rest
mass $m$ as $-(mc)^2$, so long as the theory of special
relativity is concerned. The actual admissible 
momenta only corresponds to points on the hyperboloid 
$p_\mu p^\mu=-(mc)^2$, which is a three-dimensional 
curved space. This suggests using the eight-dimensional 
vector space of $(x^\mu,p^\mu)$, or equivalently 
$(t^\mu, u^{\mu})$ with $u^{\mu}= \frac{p^\mu}{mc}$, 
the velocity four-vector in $c=1$ unit, for a Lorentz
covariant formulation. The dimensionless `momentum' 
$u^{\mu}$ is used for the conjugate variables mostly 
to match better to the group coset language. The 
value of $-(mc)^2$ though is a Casimir invariant of
the Poincar\'e symmetry which is a parameter for
characterizing a generic irreducible representation
of the symmetry \cite{T}. So, it makes good sense 
to use the momentum variables, though it really makes 
no difference when only a single particle is considered. 

The momentum or rather velocity hyperboloid 
$u_\mu u^\mu=-1$, recall $u^\mu=(\zc, \zc\zb^i)^{\!\ssc T}$,
is indeed a homogeneous space of $SO(1,3)$ corresponding 
to the coset space $SO(1,3)/SO(3)$. $SO(3)$ which keeps the 
point  $u^{\mu}=(1,0,0,0)^{\!\ssc T}$ fixed is the little 
group. A simple way to see that is to identify each point 
in the hyperboloid by the Lorentz boost that uniquely takes 
the reference point $u^{\mu}=(1,0,0,0)^{\!\ssc T}$ to it, 
hence equivalently by the coset represented by the boosts. 
Matching with the group notation as we have above, each coset 
is an $\exp(-\frac{i}{\hbar} \om^{{\ssc 0}i} J_{{\ssc 0}i}) SO(3)$. 
In fact, the coordinate for the coset $ \om^{{\ssc 0}i}=-\om^{i{\ssc 0}}$
can be identified with $-\zb^i$, for example 
from $t'^{\ssc 0}=\zc (t^{\ssc 0} + \zb_i t^i)$
giving $dt^{\ssc 0}=\bar\zb_i t^i =-\bar\om^{\ssc 0}_i t^i$.
Putting together the `phase space' as a product of
the configuration space and the momentum space, we have 
\[
ISO(1,3)/SO(1,3) \times SO(1,3)/SO(3) \;,
\]
which is mathematically exactly $ISO(1,3)/SO(3)$. 
We cannnot use it as the actual phase space in the
Hamiltonian formulation of the particle dynamics,
which has to have coordinates in conjugate pairs. 
Note that no parameter in the full Poincar\'e group 
can correspond to $u^{\ssc 0}$ and $\zb^i$ cannot 
be part of a four-vector. But there is no harm using 
the redundant coordinates $u^\mu$ to describe points
in the velocity hyperboloid. That is mathematically
a natural embedding of the velocity hyperboloid into 
the Minkowski four-vector velocity space $\mathfrak{M}_v$. 

Let us write down the explicit infinitesimal action
of $SO(1,3)$ on $SO(1,3)/SO(3)$. Note that the
translations generated by $E_\mu$ in the Poincar\'e
group do not act on the velocity four-vector 
$u^{\mu}$.  The action hence can be seen as the
full action of the  Poincar\'e group. Obviously, we 
have simply $d u^{\mu} = -\bar\om^\mu_{\,\nu}  u^{\nu}$. 
Rewriting that by taking out a $\zc=u^{\ssc 0}$ factor, 
we have 
\bea\label{db}
 d \zb^i + \zb^i\frac{d\zc}{\zc} 
= - \bar\om^i_j \zb^j + \bar\zb^i \;,
\eea
and $\frac{d\zc}{\zc}={\bar\zb_k \zb^k}$. The latter
as the extra term in the $d \zb^i$ expression shows 
the complication of the description in terms of the 
coset coordinates $\zb^i$ or $\om^{{\ssc 0}i}$ versus 
the simple picture in terms of $u^{\mu}$.

\section{Special Relativity as a Theory of Hamiltonian Dynamics}
The Hamiltonian formulation of a dynamical theory 
is a powerful one which is also particularly good for 
a symmetry theoretical formalism. Here, we consider
a coset space of the relativity symmetry group as the
particle phase space, one bearing the geometric structure
of a so-called symplectic space. The structure can be
seen as given by the existence of a Poisson bracket
as a antisymmetry bilinear structure on the algebra of
differentiable functions $F$ on the space to be given
under local coordinates $z^n$ as
\[
\{ F(z^n), F'(z^n) \} =
\Omega^{mn} \frac{\partial F}{\partial z^m} \frac{\partial F'}{\partial z^n} \;,
\qquad
\Omega^{mn} = -\Omega^{nm} \;,
\;\; \det\Omega =1 \;.
\]
In terms of canonical coordinates, for example the position
and momentum of a single (`nonrelativistic') Newtonian particle, we have
\[
\{ F(x^i, p^i), F'(x^i, p^i) \} =
\delta^{ij} \left( \frac{\partial F}{\partial x^i} \frac{\partial F'}{\partial p^j} 
- \frac{\partial F}{\partial p^j} \frac{\partial F'}{\partial x^i} \right)\;.
\]
General Hamiltonian equation of motion for any observable 
$F(z^n)$ is given by
\bea
\frac{d}{dt} F(z^n) = \{ F(z^n), {{H}}_t (z^n) \}  \;,
\eea
where ${{H}}_t (z^n)$ is the physical Hamiltonian
as the energy function on the phase space, which for case
of $F$ being $x^i$ or $p^i$ reduces to
\[
\frac{d}{dt} x^i =  \frac{\partial  {H}_t}{\partial p^i} \;, \qquad
\frac{d}{dt} p^i =  -\frac{\partial  {H}_t}{\partial x^i} \;.
\]
Note that the configuration/position variables $x^i$ 
and momentum variables $p^i$ are to be considered
the basic independent variables while the Newtonian
particle momentum being mass times velocity is to 
be retrieved from the equations of motion for the 
standard case with the $p^i$ dependent part of 
${{H}}_t$ being  $p_ip^i/2m$.

\subsection{Dynamics as Symmetry Transformations}
The key lesson here is to appreciate that the phase
space (symplectic) geometric structure guarantees 
that for any generic Hamiltonian function
${\mathcal{H}}_s$, points on the phase space 
having the same value for the function lie on a
curve of the Hamiltonian flow characterized by the
monotonically increasing real parameter $s$ 
on which any observables $F(z^n)$ satisfy the
equation
\bea\label{hs}
\frac{d}{ds} F(z^n) = \{ F(z^n), {\mathcal{H}}_s (z^n) \}  \;.
\eea
The equation of motion for the usual case is simply 
the case for ${\mathcal{H}}_t$, {\em i.e.} time evolution.
Such a physical Hamiltonian can have more than one
choice, so long as the evolution parameter is essentially 
a measure of time.  The class of Hamiltonian flows 
each generated by a Hamiltonian function having 
a vanishing Poisson bracket with the physical Hamiltonian 
function are then the symmetries of the corresponding 
physical system and the Hamiltonian functions the
related conserved quantities. In fact, a Hamiltonian 
flow is the one-parameter group of symmetry
transformations with ${\mathcal{H}}_s$ the generator 
function. We have the Hamiltonian vector field
\bea\label{Xs}
X_s = - \{  {\mathcal{H}}_s (z^n), \cdot \} 
\eea
as a differential operator being the generator and the
collection of such $X_s$ being a representation of 
the basis vectors of the symmetry Lie algebra. Hence, 
we have
\bea\label{dXs}
 \frac{dF}{ds} =  X_s (F) \;.
\eea

The structure works at least for any theory of particle
dynamics with any background relativity symmetry 
including for examples Newtonian and Einsteinian ones 
of our focus here as well as quantum mechanics.
Mathematics for the latter case is quite a bit more
involved and in many ways more natural and beautiful
from the symmetry point of view. Interested readers
are referred to Ref.\cite{070,087,094}.

\subsection{Particle Dynamics of Special Relativity}
For the phase space formulation of particle dynamics of 
special relativity, we can have a picture of the particle 
phase space as the coset space
$\mathfrak{P}:=ISO(1,3)/ T_{\!\ssc H}\times SO(3)$
with canonical coordinates $(t^k,u^k)$. The standard 
Hamilton's equations in our canonical coordinates are
\bea &&
\frac{dt_i}{dt}=\frac{\partial {\mathcal{H}_t} (t^k,u^{k})}{\partial u^{i}}\;,
\qquad
\frac{du_{i}}{dt}=-\frac{\partial {\mathcal{H}_t}(t^k,u^{k})}{\partial t^i}\;,
\eea
where Hamiltonian function 
${\mathcal{H}_t}(t^k,u^{k})=\sqrt{1+u_{k}u^{k}}$, which
is basically energy per unit mass in the dimensionless 
velocity unit ($mc^2 {\mathcal{H}_t}= c \sqrt{m^2c^2+p_{k}p^{k}}=c \, p^{\ssc 0}$).
The equations are only special cases of Eq.(\ref{hs}).
Note that the first equation really gives 
$\frac{dt_i}{dt}= \frac{u_i}{\sqrt{1+u_{k}u^{k}}}=  \zb_i$
as ${\sqrt{1+u_{k}u^{k}}}=\zc$, and the second $\frac{du_{i}}{dt}=0$.
For the extended phase space
$\mathfrak{P}_e:= \mathfrak{M}\times \mathfrak{M}_v$, 
 with canonical coordinates $(t^\mu,u ^\mu)$, we have
\bea &&
\frac{dt_\mu}{d\zeta}=\frac{\partial \tilde{\mathcal{H}}_\zeta (t^\nu,u ^\nu)}{\partial u^\mu}\;,
\qquad
\frac{du_\mu}{d\zeta}=-\frac{\partial \tilde{\mathcal{H}}_\zeta (t^\nu,u ^\nu)}{\partial t^\mu}\;,
\eea
with the extended Hamiltonian 
$\tilde{\mathcal{H}}_\zeta(t^\nu,u ^\nu)={\mathcal{H}_t} -u^{\ssc 0}$
giving, besides the same results as from ${\mathcal{H}_t}$
above, $\frac{du_{\ssc 0}}{d\zeta}=0$ for consistency and  
$\frac{dt_{\ssc 0}}{d\zeta}=-1$, hence $\zeta$ 
as essentially the coordinate time $t^{\ssc 0} \equiv t$, 
and the same dynamics \cite{J}. Alternatively, we can have
a covariant description with the proper time evolution
\bea &&
\frac{dt_\mu}{d\tau}=\frac{\partial \tilde{\mathcal{H}}_\tau (t^\nu,u ^\nu)}{\partial u^\mu} = u_\mu\;,
\qquad
\frac{du_\mu}{d\tau}=-\frac{\partial \tilde{\mathcal{H}}_\tau (t^\nu,u ^\nu)}{\partial t^\mu} =0 \;,
\eea
where $\tilde{\mathcal{H}}_\tau (t^\nu,u ^\nu)= \frac{u_\nu u^\nu}{2}$.
All formulations have equations of the form (\ref{hs}).
In fact, they can be seen all as special cases of the
single general equation from the symmetry of the 
symplectic manifold coordinated by $(t^\mu,u ^\mu)$.

We only write free particle dynamics here. The reason
being special relativity actually does not admit motion
under a nontrivial $x^\mu$ or $t^\mu$ dependent 
potential without upsetting $u_\mu u^\mu=-1$. 
Motion under gauge field, like electromagnetic field,
modifies the nature of the conjugate momentum
and the story is somewhat different.

\subsection{Hamiltonian Flows Generated by Elements of the Poincar\'e Symmetry}
We first look at the $(t^\mu, u^\mu)$ phase space 
picture. With the canonical coordinates, we have
from Eq.(\ref{Xs})
\bea &&
d t_\mu = \bar{s} \frac{\partial \tilde{\mathcal{H}}_{s}}{\partial u^\mu} \;,
\qquad
d u_\mu = -\bar{s} \frac{\partial \tilde{\mathcal{H}}_{s}}{\partial t^\mu} \;,
\eea
where $\bar{s}=ds$ is the infinitesimal parameter in
line with the notation of our coset descriptions above. 
We can see that the canonical transformations given 
by the equations for the generators of the Poincar\'e 
symmetry exactly agree with our coset picture above. 
For $\tilde{\mathcal{H}}_{\omega^{\mu\nu}} = t_\mu u_\nu - t_\nu u_\mu$,
we have $dt^\rho = -\delta^\rho_\mu \bar\omega^{\mu}_{\,\nu} t^\nu
  + \delta^\rho_\nu \bar\omega_{\mu}^{\,\nu} t^\mu$ 
and  $du^\rho = - \delta^\rho_\mu \bar\omega^{\mu}_{\,\nu} u^\nu
  + \delta^\rho_\nu \bar\omega_{\mu}^{\,\nu} u^\mu$,
while for $\tilde{\mathcal{H}}_{b^{\mu}} =  u_\mu $, 
we have $dt^\rho = \delta^{\rho}_\mu\bar{b}^\mu$, $du^\rho =0$
--- note that here we are talking about specific 
$\tilde{\mathcal{H}}_s$ functions with specific
infinitesimal parameters $\bar{s}$ on specific phase
space variables and there is no summation over any of 
the indices involved in the expressions.

The ten Hamiltonian functions
$\tilde{\mathcal{H}}_{\omega^{\mu\nu}}$ and 
$\tilde{\mathcal{H}}_{b^{\mu}}$ combined together
gives a full realization of the action of the
Poincar\'e symmetry as transformations on the
covariant phase space of $(t^\mu, u^\mu)$. One can
check that with the Poisson bracket as the Lie
bracket, they span a Lie algebra:
\bea &&
\{ \tilde{\mathcal{H}}_{\omega^{\mu\nu}}, \tilde{\mathcal{H}}_{\omega^{\lambda\rho}} \}_{\!\ssc 4}
= -(  \eta_{\nu\lambda} \tilde{\mathcal{H}}_{\omega^{\mu\rho}} 
 -  \eta_{\mu\lambda} \tilde{\mathcal{H}}_{\omega^{\nu\rho}} 
 + \eta_{\mu\rho} \tilde{\mathcal{H}}_{\omega^{\nu\lambda}} 
 -  \eta_{\nu\rho} \tilde{\mathcal{H}}_{\omega^{\mu\lambda}} ) \;,
\sea
\{ \tilde{\mathcal{H}}_{\omega^{\mu\nu}},\tilde{\mathcal{H}}_{b^{\rho}}\}_{\!\ssc 4}
= -(  \eta_{\nu\rho} \tilde{\mathcal{H}}_{b^{\mu}} - \eta_{\mu\rho} \tilde{\mathcal{H}}_{b^{\nu}} )\;,
\qquad
\{\tilde{\mathcal{H}}_{b^{\mu}},\tilde{\mathcal{H}}_{b^{\nu}}\}_{\!\ssc 4} =0 \;,
\eea
where we have explicitly 
\[
\{\tilde{\mathcal{H}}_{\ssc 1},\tilde{\mathcal{H}}_{\ssc 2}\}_{\!\ssc 4}
 = \eta^{\mu\nu} \left( \frac{\partial \tilde{\mathcal{H}}_{\ssc 1}}{\partial t^\mu} \frac{\partial \tilde{\mathcal{H}}_{\ssc 2}}{\partial u^\nu}
- \frac{\partial \tilde{\mathcal{H}}_{\ssc 1}}{\partial u^\nu} \frac{\partial \tilde{\mathcal{H}}_{\ssc 2}}{\partial t^\mu} \right) \;.
\]
Matching $\tilde{\mathcal{H}}_{\omega^{\mu\nu}}$ to
$i\hbar J_{\mu\nu}$ and $\tilde{\mathcal{H}}_{b^{\mu}}$ 
to $i\hbar E_{\mu}$ we can see that the Lie algebra
is that of the Poincar\'e symmetry given by Eq.(\ref{PS}).
In fact, it is a representation of the symmetry on the
space of functions of the phase space variables.
If the phase space $\mathfrak{P}$ is taken, however,
we can have only as Hamiltonian functions
${\mathcal{H}}_{\omega^{ij}}$ and ${\mathcal{H}}_{b^{i}}$,
with identical expressions to $\tilde{\mathcal{H}}_{\omega^{ij}}$ 
and $\tilde{\mathcal{H}}_{b^{i}}$, illustrating only 
the $ISO(3)$ symmetry of translations and rotations,
with the Lie product
\[
\{ {\mathcal{H}}_{\ssc 1},{\mathcal{H}}_{\ssc 2}\}_{\!\ssc 3}
 = \eta^{ij} \left( \frac{\partial {\mathcal{H}}_{\ssc 1}}{\partial t^i} \frac{\partial {\mathcal{H}}_{\ssc 2}}{\partial u^j}
- \frac{\partial {\mathcal{H}}_{\ssc 1}}{\partial u^j} \frac{\partial {\mathcal{H}}_{\ssc 2}}{\partial t^i} \right) \;.
\]
The time translation symmetry can be added with 
${\mathcal{H}}_{t}$ given above, which has the right
vanishing Lie product as 
$\{ {\mathcal{H}}_{\omega^{ij}},{\mathcal{H}}_t \}_{\!\ssc 3}=0$ 
and $\{ {\mathcal{H}}_{b^{i}},{\mathcal{H}}_t \}_{\!\ssc 3}=0$.
Not being able to have the boosts as Hamiltonian
transformations is one of the short-coming of
not using the covariant phase space.

\section{Contractions as Approximations of Physical Theories}
With an understanding of how the principle of relativity 
informs our notion of physical spacetime and the theory
of particle dynamics  behind us, we can move on to the 
important connection this language provides us between 
different  theories from the relativity symmetry perspective. 
Broadly speaking this can be put as: it is commonplace 
to find phrases like ``Newtonian physics arises from
special relativity when $c\to\infty$'' and we will place 
such comments on a firm mathematical foundation
within the relativity theoretical symmetry setting.

\subsection{A Crash Course on Symmetry Contractions}
Imagine we are standing on a perfectly spherical, uninhabited 
planetary body\footnote{
This example, and indeed the entire examination of 
contractions found here, is strongly influenced by 
the wonderful discussion found in \cite{Gil}.}. 
The  transformation that arise as symmetries 
of said body are  nothing more than the $SO(3)$
group elements as rotations about the center.
Now consider what we can say if this body began 
to rapidly expand without limit. It is intuitively 
clear that as the radius of the sphere becomes 
larger and larger, making the surface of the sphere 
more and more flat, the symmetries of this body 
should be approaching, in some sense, those of 
the Euclidean plane, {\em i.e.} $ISO(2)$.  It might 
not, however, be immediately clear how exactly 
this is encoded in the structure of the Lie algebras. 
How might one achieve such a description? It is 
ultimately this question, applied to a general Lie 
algebra $\mathfrak{g}$, that we are concerned 
with in this section. The notion of a 
\textit{contraction} is precisely the answer 
we are looking for.

In particular, we will focus on the simplest form of 
contractions: the so-called \textit{In\"{o}n\"{u}-Wigner 
contractions} \cite{IW}. The setup is as follows: 
consider a Lie algebra $\mathfrak{g}$ with a 
decomposition $\mathfrak{g}=\mathfrak{h}+\mathfrak{p}$, 
where $\mathfrak{h}$ is an $n$-dimensional subalgebra 
and $\mathfrak{p}$ the complementary $m$-dimensional 
vector subspace.  In terms of our example above, the idea
 is that we collect the portion of the symmetries that do 
not change in the limit (which are the rotations around 
the vertical axis through where we stand on the planet, 
for the example at hand) and call their Lie algebra 
$\mathfrak{h}$. The rest, or the span of the
independent generators  is  $\mathfrak{p}$. Then we 
can form a one-parameter sequence of base changes, 
corresponding directly to the change in scale of the 
physical system, of the form
\[\left( \begin{array}{c}
\mathfrak{h} \\ 
\mathfrak{p}'
\end{array}  \right)=\left( \begin{array}{cc}
I_n & 0 \\ 
0 & \frac{1}{R}  I_m
\end{array}  \right)\left( \begin{array}{c}
\mathfrak{h} \\ 
\mathfrak{p}
\end{array}  \right)\]
for any nonzero value of $R$ here taken conveniently 
as positive. For any finite $R$, the Lie algebra hence our
symmetry is not changed. In the $R \to \infty$ limit, 
however,we obtain the contracted algebra 
$\mathfrak{g}'=\mathfrak{h}\oplus\mathfrak{p}'$.
Note that, although the change of basis matrix is 
singular in the limit, the commutation relations 
still make sense:
\begin{align*}
[\mathfrak{h},\mathfrak{h}] = [\mathfrak{h},\mathfrak{h}]
   & \subseteq \mathfrak{h}
  \qquad   \qquad    \qquad\qquad
   \xrightarrow{\quad R \to \infty \quad}  \quad \mathfrak{h} \;, \\
[\mathfrak{h}',\mathfrak{p}'] = \frac{1}{R}  [\mathfrak{h},\mathfrak{p}]
 &  \subseteq \frac{1}{R} (\mathfrak{h}+\mathfrak{p})
   =\frac{1}{R} \mathfrak{h}+\mathfrak{p}' 
  \quad  \xrightarrow{\quad R \to \infty\quad} \quad  \mathfrak{p}'\;, \\
[\mathfrak{p}',\mathfrak{p}'] = \frac{1}{R^2} [\mathfrak{p},\mathfrak{p}]
  & \subseteq \frac{1}{R^2} (\mathfrak{h}+\mathfrak{p})
  = \frac{1}{R^2} \mathfrak{h}+\frac{1}{R} \mathfrak{p}' 
  \xrightarrow{\quad R \to \infty \quad}  \quad 0 \;.
\end{align*}
Though the vector space is the same, the Lie
products, or commutators, change. $\mathfrak{p}$
is in general not even a subalgebra of $\mathfrak{g}$.
$\mathfrak{p}'$ is however an Abelian subalgebra 
of $\mathfrak{g}'$ and is an invariant one.

Take the explicit example we have. The Lie algebra 
$\mathfrak{so}(3)$ for the group $SO(3)$ is given by
 the commutation relations
\[
[J_x,J_y]= i\hbar  J_z, \quad[J_y,J_z]=i\hbar J_x, 
\quad \textnormal{and} \quad [J_z,J_x]=i\hbar J_y \;.
\]
Under the rescaling  $P_x=\tfrac{1}{R}J_x$,
$P_y=\tfrac{1}{R}J_y$, and $J_z$ as the generator of
$\mathfrak{h}$ is not changed (taking a coordinate 
system with where we stand as the on the positive 
$z$-axis), the commutators become
\begin{align*}
&[P_x,P_y]=\frac{1}{R^2}[J_x,J_y]
 =\frac{i\hbar}{R^2}{J}_z \to 0 \;, \\
&[{J}_z,P_x]=\frac{1}{R}[J_z,J_x]= i\hbar P_y \;, \\
&[{J}_z,P_y]=\frac{1}{R}[J_z,J_y]=- i\hbar P_x \;,
\end{align*}
in the limit as $R\rightarrow\infty$. Therefore, we recover 
precisely commutation relations of the Lie algebra\footnote{
Our notation is such that it has a nice matching 
to the Poincar\'e symmetry ones used above, 
with the identification of $J_x$,  $J_y$,
$J_z$, $P_x$ and $P_y$ as $J_{\!\ssc 23}$, $J_{\!\ssc 31}$,
$J_{\!\ssc 12}$, $E_{\!\ssc 1}$ and $E_{\!\ssc 2}$,
respectively. 
} 
$\mathfrak{iso}(2)$. From the physical 
geometric perspective, we see that what is really 
happening in the limit is that the ratio of the 
characteristic distance scales we have chosen, like 
the length of our foot step or the distance we can 
travel and that of the radius, is becoming zero. The
radius $R$ is effectively infinity to us if we can 
only manage to explore a distance tiny in comparison. 
The planet is as good as flat to us then, though it
is only an approximation.

\subsection{The Poincar\'e to Galilean Symmetry Contraction}
Our starting point for describing the transition from 
Einsteinian relativity to Galilean relativity is the 
following natural choice of a contraction of the 
Poincar\'e algebra to the Galilean algebra. Moreover, 
we will see that this takes Minkowski spacetime, viewed 
as a coset space of $ISO(1,3)$, to ordinary Newtonian 
space-time, viewed as a coset space of $G(3)$. Actually,
it goes all the way to take the full dynamical theory
as given by the symplectic geometry of the phase space
as a representation space from that of special relativity
to the Newtonian one.

The contraction is performed via the new generators 
$K_i= \frac{1}{c} J_{i{\ssc 0}}$ and  $P_i = \frac{1}{c} E_i$, 
keeping $J_{ij}$ and $E_{\!\ssc 0}$  is renamed $-H$.  
Then we have
\bea &&
[J_{ij}, J_{hk}] = - i \hbar ( \delta_{jh}  J_{ik}  - \delta_{ih}  J_{jk}
   + \delta_{ik} J_{jh}- \delta_{jk} J_{ih})\;,
\nonumber \\ &&
[J_{jk}, H] =0 \;,
\eea
which is the subalgebra that is not rescaled 
($\eta_{ij}=\delta_{ij}$).
As for the other commutators, we have
\bea &&
[J_{ij}, K_k]= \frac{1}{c} [ J_{ij}, J_{k{\ssc 0}} ]
  = - i \hbar ( \delta_{jk}  K_i -\delta_{ik} K_j ) \;,
\nonumber \\ &&
[K_i, K_j] =\frac{1}{c^2} [J_{i{\ssc 0}}, J_{j{\ssc 0}}]
  = -i \hbar\frac{1}{c^2} J_{ij} \;,
\nonumber \\ &&
[J_{ij}, P_k ] = \frac{1}{c} [J_{ij}, E_k ] 
 = -i  \hbar (\delta_{jk} P_i -\delta_{ik}  P_j) \;,
\nonumber \\ &&
[J_{ij}, H] = 0 \;,
\nonumber \\ &&
[K_i, P_j] = \frac{1}{c^2} [ J_{i\ssc 0}, E_{j} ]
  = -i \hbar\frac{1}{c^2}  \eta_{ij} H  \;,
\nonumber \\ &&
[K_i, H] =  -\frac{1}{c} [J_{i{\ssc 0}}, E_{\ssc 0}]
  = -i \hbar P_i \;,
\nonumber \\ &&
[H, P_i] = -\frac{1}{c}[E_{\ssc 0}, E_i]=0 \;,
\nonumber \\ &&
[P_i, P_j]= \frac{1}{c^2}[E_i, E_j]=0 \;.
\eea
When we take the $c\to\infty$ limit,  we have
$[K_i, K_j] =0$ and $[K_i, P_j] =0$. That is, we recover 
the Galilean symmetry algebra. Note that we need
the  $\tfrac{1}{c}$ factor in $K_i= \tfrac{1}{c} J_{i{\ssc 0}}$ 
in order to get $[K_i, K_j] =0$, hence, Lorentz boosts
becoming commutating Galilean boosts.  Moreover,
 this will give $[K_i, P_j]=0$ as well if we simply take
$P_i=E_i$.  However, this will also yield $[K_i, H] =0$
in the contraction limit which cannot be the Galilean
symmetry. By taking $P_i = \frac{1}{c} E_i$ though, one 
can see that this saves $[K_i, H] =-i \hbar P_i $, as 
needed. This is actually precisely the reason we wanted 
to start with $E^\mu$, instead of $P^\mu$! Indeed, the 
momentum $P_i$ are not the generators of the Poincar\'e 
algebra we started with before the introduction of the 
nontrivial factor of $c$. 

The mathematical formulation of the contraction above
can also be understood from a geometric picture. It 
is about an approximation when the relevant velocities
of particle motion have magnitudes small relative to 
the speed of light $c$, {\em i.e.} $\beta^i <\!\!< 1$.
The velocity space for particle motion under special
relativity is the three-dimensional hyperboloid of
`radius' $c$ -- the four-velocity $cu^\mu$ is a 
timelike vector of magnitude $c$. When we are only
looking at a small region around zero motion of
$u^\mu=(1,0,0,0)^{\!\ssc T}$, the velocity space
seems to be flat, like the {\em Euclidean} space
of Newtonian three-velocity $v^i$, and the boosts
as commuting velocity translations.

\subsection{Retrieving Newtonian Space-Time from Minkowski Spacetime}
Now we can parse the changes in the Minkowski spacetime 
coordinates $t^\mu$, as a representation, under the 
contraction. First of all, we have to write our algebra 
elements in terms of these new generators, in order to 
paint a coherent picture. We have
\bea
-\frac{i}{\hbar}\left(\frac{1}{2}\omega^{\mu\nu} J_{\mu\nu}+ b^\mu E_\mu\right)
&=&-\frac{i}{\hbar}\left(\frac{1}{2}\omega^{ij} J_{ij}+ b^{\ssc 0} E_{\ssc 0} 
+\omega^{{\ssc 0}i} J_{{\ssc 0}i}+ b^{i} E_{i} \right)
\nonumber \\  
&=&-\frac{i}{\hbar}\left(\frac{1}{2}\omega^{ij} J_{ij}+ b^{\ssc 0} E_{\ssc 0} 
+c\,\omega^{i{\ssc 0}} K_i+ c\,b^{i} P_{i} \right)
\nonumber \\
&=&-\frac{i}{\hbar}\left(\frac{1}{2}\omega^{ij} J_{ij}+ b^{\ssc 0} E_{\ssc 0} 
+v^i K_i+ a^{i} P_{i} \right),
\eea
where $v^i= c \,\omega^{i{\ssc 0}}$ and $a^i =c \,b^i$ are the new
parameters for the boosts and spatial translations
(i.e. the $x^i$ translations). 
The representation for the algebra is given by
\bea
\left(\begin{array}{c}
dt \\ dx^i =c \, dt^i\\ 0
\end{array}\right)
&\equiv&
\left(\begin{array}{ccc}
0 & -\frac{1}{c} \bar\omega^{\ssc 0}_{\,j} & \bar{b} \\
 -c\,\bar\omega^i_{\,\ssc 0} & -\bar\omega^i_{\,j}  & \bar{a}^i =c\,\bar{b}^i \\
0 & 0 & 0
\end{array}\right)
\left(\begin{array}{c}
t \\ x^i= c \, t^i\\ 1
\end{array}\right)
= \left(\begin{array}{c}
 -\frac{1}{c} \bar\omega^{\ssc 0}_{\,j} x^j+ \bar{b} \\
  -c\,\bar\omega^i_{\,\ssc 0} t - \bar\omega^i_{\,j} x^j + \bar{a}^i \\ 0
\end{array}\right) 
\nonumber \\
&=&
\left(\begin{array}{ccc}
0 & \frac{1}{c^2} \bar{v}_j & \bar{b} \\
\bar{v}^i & -\bar\omega^i_{\,j}  & \bar{a}^i \\
0 & 0 & 0
\end{array}\right)
\left(\begin{array}{c}
t \\ x^i \\ 1
\end{array}\right)
= \left(\begin{array}{c}
 \frac{1}{c^2}  \bar{v}_j x^j+ \bar{b} \\
  \bar{v}^i t - \bar\omega^i_{\,j} x^j + \bar{a}^i \\ 0
\end{array}\right) 
\;.
\eea
where we have used
\[
c\,\bar\omega^i_{\,\ssc 0} 
=   -c \, \bar\omega^{i{\ssc 0} } = - c\, \bar\zb^{i} = - \bar{v}^i \;,
\qquad
 {\bar\omega^{\ssc 0}_{\,j}} =  {\bar\omega^{{\ssc 0}j}} =-\frac{1}{c}  v_j \;.
\]
Lastly, we take the limit $c\to\infty$ and get
\bea
\left(\begin{array}{c}
dt \\ dx^i \\ 0
\end{array}\right)
=\left(\begin{array}{ccc}
0 & 0 & \bar{b} \\
\bar{v}^i & -\bar\omega^i_{\,j}  & \bar{a}^i \\
0 & 0 & 0
\end{array}\right)
\left(\begin{array}{c}
t \\ x^i \\ 1
\end{array}\right)
= \left(\begin{array}{c}
  \bar{b} \\
  \bar{v}^i t - \bar\omega^i_{\,j} x^j + \bar{a}^i \\ 0
\end{array}\right) 
\;.
\eea
The group of finite transformations can be written in the form
\bea
\left(\begin{array}{c}
t' \\ x'^i \\ 1
\end{array}\right)
=\left(\begin{array}{ccc}
1 & 0 & B \\
V^i & R^i_{\,j}  & A^i \\
0 & 0 & 1
\end{array}\right)
\left(\begin{array}{c}
t \\ x^i \\ 1
\end{array}\right)
= \left(\begin{array}{c}
  t+B \\
  V^i t + R^i_{\,j} x^j + A^i \\ 1
\end{array}\right) 
\;.
\eea
Newtonian space-time with transformations under a generic
element in the Galilean group has been retrieved. Now we 
can see that the  Newtonian space-time `points' can be 
described by the coset 
\[
\left(\begin{array}{ccc}
1 & 0 & t \\
V^i & R^i_{\,j}  & x^i \\
0 & 0 & 1
\end{array}\right)
=\left(\begin{array}{ccc}
1 & 0 & t \\
0 & \delta^i_k  & x^i \\
0 & 0 & 1
\end{array}\right)
\left(\begin{array}{ccc}
1 & 0 & 0 \\
V^k & R^k_{\,j}  & 0 \\
0 & 0 & 1
\end{array}\right)\;,
\]
as
\[
\left(\begin{array}{ccc}
1 & 0 & t \\
V^i & R^i_{\,j}  & x^i \\
0 & 0 & 1
\end{array}\right)
\left(\begin{array}{c}
0 \\ 0 \\ 1
\end{array}\right)
= \left(\begin{array}{c}
t \\ x^i \\ 1
\end{array}\right)\;.
\]
Indeed, the matrix expressed as that product of two
is exactly in the form of the first matrix representing
a particular element $\exp\!\big(-\frac{i}{\hbar}(tH +x^iP_i)\big)$ 
of pure translations multiply to any element with the
rotations and Galilean boosts, as translations on the
space of Newtonian velocity, only, hence any element of 
the coset $\exp\!\big(-\frac{i}{\hbar}(tH +x^iP_i)\big) ISO_v(3)$. 
The Newtonian space-time as a coset space is given by
$G(3)/ISO_v(3)$, and the $ISO_v(3)$ subgroup is 
exactly the result of the contraction from $SO(1,3)$,
{\em i.e.} we have
\[
ISO(1,3)/SO(1,3) \longrightarrow G(3)/ISO_v(3) \;.
\]
The infinitesimal action of the $G(3)$ group on the
coset here obtained from the contraction may also
be obtained directly from first principle. The
simpler commutation relations actually make the 
calculation easier.

In Einstein relativity, spacetime should be described by 
coordinates with the same units. The natural units are 
given by the $c=1$ units, which identifies each spatial 
distance unit with a time unit, and vice versa.  If one
insists on using different units for the time and space 
parts, $c$ has then the unit of distance over time and 
can be written as any value in different units, like 
$\sim 3\times 10^{8}\, ms^{-1}$, or $\sim 3\times 10^{28}\, A \, yr^{-1}$,
or $\sim 3 \times 10^{-7}\, km \,ps^{-1}$,
or $\sim 10^{-26} \, Mpc \, ps^{-1}$.  The exact choice 
of units is arbitrary. The structure of the physical  
theory is independent of that. Hence, any finite value 
of $c$ describes the same symmetry represented by 
spacetime coordinates in different units. The $c\to\infty$
limit is different. Infinity is infinity in any units, 
and the algebra becomes the contracted one, which is 
to say that the relativity symmetry becomes Galilean.
The latter is practical as an approximation for physics 
at velocity much less than $c$. Pictured in the Minkowski 
spacetime, such lines of motion hardly deviate from the
time axis, giving the idea of the Newtonian absolute
time.  The relativity symmetry contraction picture gives 
a coherent description of all aspects of that approximate 
theory, including the dynamics to which we will turn below.

\subsection{Hamiltonian Transformations and Particle Dynamics at the Newtonian Limit}
Turning to the phase space pictures, we have
already $dt^\mu = -\bar\om^\mu_\nu t^\nu +\bar{t}^\mu$
giving at the contraction limit $dt=\bar{t}$ and
$dx^i= \bar{v}^i t - \om^i_j x^j +\bar{x}^i$. 
Similarly, we can see that  
\bea &&
du^{\ssc 0}= -\bar\om^{\ssc 0}_i u^i = \bar\zb_i u^i
 \quad\Longrightarrow \quad d\zc = \frac{\bar{v}_i v^i \zc}{c^2} \to 0 \;,
\sea
du^i = -\bar\om^i_\nu u^\nu 
 \quad\Longrightarrow\quad dv^i = - v^i\frac{d\zc}{\zc} -\bar\om^i_j v^j + \bar{v}^i
  \to -\bar\om^i_j v^j + \bar{v}^i \;,
\eea
where we have used Eq.(\ref{db}). The phase space
$\mathfrak{P}$ at the contraction limit should be
described with coordinates $(x^i,v^i)$. The coset
space of $ISO(1,3)/SO(3)$ or $\mathfrak{P}_e$ with 
$(t,x^i,v^i)$ as $\zc \to 1$ can no longer have $t$ as a 
meaningful coordinate. We have then only one
sensible phase space as the Newtonian one here
with $(x^i,v^i)$.

To look at the Hamiltonian symmetry flows or the
dynamics at the contraction limit, the notation 
of the Hamiltonian vector field is convenient. On
$\mathfrak{P}$ with $\{\cdot,\cdot \}_{\ssc 3}$
we have 
\bea
 X_s^{\ssc (3)} &=& - \{ {\mathcal{H}}_s(t^i,u^i),\cdot\}_{\!\ssc 3}
 = - \eta^{ij} \left( \frac{\partial {\mathcal{H}}_s(t^i,u^i)}{\partial t^i} \frac{\partial}{\partial u^j}
- \frac{\partial {\mathcal{H}}_s(t^i,u^i)}{\partial u^j} \frac{\partial }{\partial t^i} \right)
\sea
 = - \delta^{ij} \left( \frac{\partial {c^2\mathcal{H}}_s(x^i,v^i)}{\partial x^i} \frac{\partial}{\partial (\zc v^j)}
- \frac{\partial {c^2\mathcal{H}}_s(x^i,v^i)}{\partial (\zc v^j)} \frac{\partial }{\partial x^i} \right)\;.
\eea
For ${\mathcal{H}}_t=\sqrt{1+u_ku^k}$ in particular, we have
$c^2 {\mathcal{H}}_t=c^2 + \frac{1}{2}\zc^2 v_kv^k + \dots$
where the terms not shown contain negative powers of $c^2$
and vanish at the $c \to \infty$ contraction limit. 
Multiply by the mass $m$ and take the expression to the
contraction limit, the first term is diverging, but is
really the constant rest mass contribution to energy,
while the finite second term is the kinetic energy
$m H_t=\frac{1}{2}m v_kv^k$. Anyway, the $c^2$ term being 
constant does not contribute to $X_t^{\ssc (3)}$, which
then reduces to 
\bea
 X_t^{\ssc (3)} \quad \to \quad X_t
 = - \delta^{ij} \left( \frac{\partial H_t}{\partial x^i} \frac{\partial}{\partial v^j}
- \frac{\partial H_t}{\partial v^j} \frac{\partial }{\partial x^i} \right)\;.
\eea
The Hamilton's equations of motion are more directly
giving $\frac{dx_i}{dt}=v_i$ and $\frac{dv_i}{dt}=0$.
We have retrieved free particle dynamics of the
Newtonian theory, though with the mass $m$ dropped
from the description. The case with a nontrivial
potential energy $V$ can obviously be given by
taking $H_t=\frac{1}{2} v_kv^k + \frac{V}{m}$. The fact
that the case cannot be retrieved from the contraction
limit of special relativity is a limitation of the
latter which cannot describe potential interaction
other than those from gauge fields \cite{noi}.

On the Lorentz covariant phase space, we have 
\bea
 X_s^{\ssc (4)} &=& - \{ \tilde{\mathcal{H}}_s(t^\mu,u^\mu),\cdot\}_{\!\ssc 4}
 = - \{ \tilde{\mathcal{H}}_s,\cdot\}_{\!\ssc 3} 
  - \eta^{\ssc 00} \left( \frac{\partial \tilde{\mathcal{H}}_s}{\partial t^{\ssc 0}} \frac{\partial}{\partial u^{\ssc 0}}
- \frac{\partial \tilde{\mathcal{H}}_s}{\partial u^{\ssc 0}} \frac{\partial }{\partial t^{\ssc 0}} \right)\;.
\eea
This, together with the above, shows that for $c \to \infty$
\bea
 X_\zeta^{\ssc (4)} = X_t^{\ssc (3)} + \frac{\partial }{\partial t^{\ssc 0}} 
 \quad \to \quad X_t  + \frac{\partial }{\partial t} \;,
\eea
giving the same dynamics. Similarly, we have 
$X_\tau^{\ssc (4)}$ giving the same limit, as 
$c^2 \tilde{\mathcal{H}}_\tau = H_t + \frac{c^2\zc^2}{2}
  \to H_t + \frac{c^2}{2}$. The exact limits of the
$X_s^{\ssc (4)}$ are generally vector fields on the 
space of $(t,x^i,v^i)$ though. The space can be seen 
as an extension of the Newtonian phase space with 
the time coordinate, and a Poisson bracket defined 
independent of the latter. The true Hamiltonian 
vector field as a vector field on the phase space 
should have the $t$ part dropped from consideration, 
like $X_t  + \frac{\partial }{\partial t}$ projected
onto $X_t$.

Further extending the analysis to the Hamiltonian
functions  ${\mathcal{H}}_{\om^{ij}}=\tilde{\mathcal{H}}_{\om^{ij}}$, 
${\mathcal{H}}_{b^{i}}=\tilde{\mathcal{H}}_{b^{i}}$,
${\mathcal{H}}_t$, $\tilde{\mathcal{H}}_{\om^{\ssc i0}}$, 
and $\tilde{\mathcal{H}}_{b^{\ssc 0}}$, one can retrieve
\bea &&
\{ H_{\omega^{ij}}, H_{\omega^{hk}} \}
= -(  \delta_{jh} H_{\omega^{ik}} 
 -  \delta_{ih} H_{\omega^{jk}} 
 + \delta_{ik} H_{\omega^{jh}} 
 -  \delta_{jk} H_{\omega^{ih}} ) \;,
\sea
\{ H_{\omega^{ij}},H_{v^{k}}\}
= -(  \delta_{jk} H_{v^{i}} - \delta_{ik} H_{v^{j}} )\;,
\qquad
\{H_{v^{i}},H_{v^{j}}\} =0 \;,
\sea
\{ H_{\omega^{ik}},H_{a^{k}}\}
= -(  \delta_{jk} H_{a^{i}} - \delta_{ik} H_{a^{j}} )\;,
\qquad
\{H_{a^{i}},H_{a^{j}}\} =0 \;,
\sea
\{H_{v^{i}},H_{a^{j}}\} = -\delta_{ij} \;,
\qquad
\{H_{v^{i}},H_{t}\} = - H_{a^{i}} \;,
\qquad
\{H_{t},H_{a^{i}}\} =0 \;,
\sea
\{ H_{\omega^{ij}}, H_{t} \} =0 \;,
\eea
with $H_{\omega^{ij}}  = x_i v_j -x_j v_i$, 
$H_{v^{i}} =  -x_i$, and $H_{a^{i}}= v_i$. We have 
already looked at $H_t$ from $c^2 {\mathcal{H}}_t$. 
$\tilde{\mathcal{H}}_{b^{\ssc 0}}= u_{\ssc 0}$ can
of course be rewritten as $ - {\mathcal{H}}_t$,
which is in line with the Galilean generator $H$ as $-E_{\ssc 0}$. 
Note that the set of Hamiltonian vector fields as differential
operators serve as a representation of the Lie algebra 
generators, as infinitesimal transformations on the phase space,
with their commutators as the Lie product/brackets. While
the corresponding Hamiltonian functions are usually talked
about as generating function or even generators for the
Hamiltonian flows on the phase space,  they are really elements 
of the observable algebra as the corresponding representation
of the universal enveloping algebra or some extension of the
group algebra with the simple functional product. The Lie 
product/brackets is there represented again as the 
commutator which vanishes between all Hamiltonian functions. 
The Poisson bracket as an alternative Lie bracket on the latter 
realizes rather the Lie bracket of the representation of the $U(1)$
central extension of the Galilean symmetry  \cite{Gil,gq}.  The
 latter is essentially the relativity symmetry for the in quantum 
mechanics. In fact, the `mismatch' between the two parts, as 
$\{H_{v^{i}},H_{a^{j}}\} = -\delta_{ij}$  versus
$[X_{v^{i}},X_{a^{j}}]=[K_i,P_j]=0$  can be better understood 
from the symmetry contraction of the quantum theory, hence 
can be seen to have a quantum origin \cite{070}. 
$H_{\omega^{ij}}$ and $H_{a^{i}}= v_i$ are from
the limit of $c^2 \tilde{\mathcal{H}}_{\om^{ij}}$
and $c \tilde{\mathcal{H}}_{b^{i}}$, respectively.
We can see from the above Hamiltonian vector field
analysis that using the $(x^i,v^i)$ instead of
$(t^i,u^i)$ as canonical coordinates implies a $c^2$
factor for the matching Hamiltonian function. The
 $\frac{1}{c}$ extra factor in $H_{a^{i}}$ is from
the symmetry contraction of $P_i = \frac{1}{c} E_i$.
The a bit more complicated case is with the 
Hamiltonian generator for the Galilean boosts
$H_{v^{i}}$. We are supposed to take again 
$c \tilde{\mathcal{H}}_{\om^{0i}}$ to the 
$c \to \infty$ limit, which gives $v_i t- x_i$
and the Hamiltonian vector field as
$\frac{\partial }{\partial v^i} - t \frac{\partial }{\partial x^i}$. 
Projecting that vector field on space of $(t,x^i,v^i)$ 
to a Hamiltonian vector field on the phase space
gives only $\frac{\partial }{\partial v^i}$,
which corresponds to our Hamiltonian function
of $H_{v^{i}}=-x_i$. If $v_i t- x_i$ is naively
taken, all expressions would be the same except
$\{H_{v^{i}},H_{v^{j}}\}$ which would then be
 $-2t \delta_{ij}$.

\section{Concluding Remarks}
As we have seen above, quite a lot of information about 
our description of physical systems is actually encoded 
in the underlying relativity symmetry algebra. What we hope 
to emphasize here is that this is really a great, if not the only
correct, perspective from which one can classify a physical 
paradigm, as well as the possible extensions and approximations.

It is also important to note that this story is not unique 
to special relativity and the Newtonian limit. There is, 
indeed, an additional question motivating this note, namely,
how the symmetry perspective can be used to understand 
better quantum mechanics and its classical limit. The 
relativity symmetry contraction picture can be seen as
a way to understand the classical phase space as an
approximation to the quantum phase space \cite{070},
and even suggests a notion of a quantum model for the
physical space \cite{066}. In a broader scope, relativity
symmetry deformations has been much pursued as a probe 
to possible dynamical theories at the more fundamental
levels \cite{dsr,dsr2,BL,dsr3,dsr31}.

Concerning the classical theories with some coset spaces
serving essentially as the phase spaces for dynamical
theories under the corresponding symmetries, it should 
be mentioned that the so-called coadjoint orbits of Lie 
groups are essentially the only nontrivial mathematical 
candidates for symplectic geometries. The full structures 
of all such symplectic geometries and hence dynamical 
theories can be derived \cite{gq,GS,ims} though the detailes 
mathematics is not so easily appreciable to many students. 
Coset spaces are also the natural candidates for 
homogeneous geometric spaces.

\begin{appendices}

\section{Deformations as Probes of More Fundamental Physics}

We would expect the symmetry of a physical system to be 
robust under small perturbations, as otherwise our limited
precise in measurements would imply that we can \textit{never} 
actually correctly determine or identify the symmetry of 
a given system. Indeed, the fact that a minute perturbation 
-- too small to be detected by our best measuring apparatuses 
-- could yield a different symmetry Lie algebra than the actual 
one means that we are epistemologically blind to the underlying 
physics. As such, it makes sense to focus our attention on 
algebras that \textit{are} significantly robust under small 
perturbations. 

For a Lie algebra, a perturbation can be taken as a (small) 
modification of the structural constants. For example, taking 
the Lorentz symmetry of $SO(1,3)$ with generators at the 
standard physical units, the commutators/Lie brackets among 
the infinitesimal Lorentz boost generators is proportional to
$1/c^2$, as can be seen in the main text. Actually, for any 
finite value of $c$, the symmetry is the same mathematical
group/algebra. Again, at $1/c^2=0$, {\em i.e.} speed of light 
being infinity, it is a different symmetry, the $ISO(3)$ of
rotations and Galilean boosts. If we have not measured the
finite speed of light, we would only be able to have an
experimental lower bound for it. Confirming $1/c^2=0$ 
requires infinite precision, which can never be available.
It makes sense then to prefer the Lorentz symmetry and 
see the Galilean one as probably only an approximation at
physical velocities small compared to the yet undetermined 
large speed of light. That is more or less the argument
Minkowski had \cite{D,Min} on one could have discovered 
special relativity from mathematical thinking alone. Here 
it is only about the idea of the zero structural constant in 
the Galilean symmetry making it unstable upon perturbations,
or deformations. The physical identification of the constant 
as $1/c^2$ is not even necessary. It is straightforward 
to check that $SO(1,3)$ and $SO(4)$ are the only possible
deformations of $ISO(3)$, within the Lie group/algebra
setting. And they themselves are stable against deformation.

One can argue that we have a similar situation with the
commutator between a pair of position and momentum 
operators as generators for the Heisenberg-Weyl symmetry 
behind quantum mechanics. Actually, the zero commutator 
limit of which can be essentially identified as that between
the $K_i$ and $P_i$ generators of the Galilean symmetry,
with $K_i=mX_i$ and $m$ being the particle mass. Then,
one can also further contemplate deforming the zero
commutators among the position and momentum 
operators, all the way till reaching a stable Lie algebra, 
one that no further deformation to a different Lie algebra 
is mathematically possible \cite{dsr2}. Within the Lie
group/algebra symmetry framework, the scheme may 
suggest a bottom-up approach to construct some plausible 
more fundamental theories.

\section{A Physicist's Sketch of the Necessary Group Theory Background}
A group is the abstract mathematical description of 
a system of symmetries, or symmetry transformations.
Lie groups are continuous symmetry transformations,
like a collection of rotations through any possible real 
value of the angle. A symmetry group of a geometric 
space can be seen as a set of transformations that do 
not change the space.  Note that apparently different
groups of transformations on different spaces may be
mathematically the same group. For example, the 
group of rotations, around the origin, on a plane 
is mathematically identified with the group of 
translations along the circle as a one-dimensional
(curved) space. From the abstract mathematical point
of view, each transformation is an abstract element
of the group as a collection, which is given with a set
of conditions, the group axioms, on the not necessarily
commutative product defined between the element
to be satisfied. The latter is automatic for a Lie group
defined as below. When a group is described as 
transformations on a vector space, that is called 
a representation of the group.  Physicists often start 
formulating a group from one of its representations, 
like as symmetry transformations on a model of the 
physical space or phase space for a particle. 

For continuous symmetries, we would like to 
think about their infinitesimal counterparts. The 
mathematical description, or abstraction, is given 
by a (real) Lie algebra. It has a set of generators
$X_i$ giving a generic element as a linear combination 
$a^i X_i$ (summed over $i$) for any set of real 
numbers $a^i$. Or we talk about the $a^i$ as real
parameters, and each distinct set of values for them 
specifies an element. The Lie algebra is further
defined by having a Lie product, $[\cdot,\cdot]$,
between its elements given with 
\[
[X_j, X_k] = c_{jk}^i X_i
\]
which is antisymmetric ($[X,Y]=-[Y,X]$) and
satisfies the Jacobi identity
\[ 
[[X,Y],Z] + [[Y,Z],X] + [Z,X],Y] =0 \;.
\]
The real numbers $c_{jk}^i$, not all independent,  are 
called the structural constants of the Lie algebra. A Lie 
algebra is hence an abstract vector space, with no notion 
of vector magnitude or inner product, and the set of 
generators a basis of it. 

A generic element of the Lie group $G$ with an associated
Lie algebra $\mathfrak{g}$ can be written as $\exp(a^i X_i)$,
the formal power series, and an infinitesimal transformation 
as $\exp(\bar{a}^i X_i) =1+\bar{a}^i X_i$ for infinitesimals
$\bar{a}^i$ ($1$ being the identity transformation,
{\em i.e.} no transformation). Note that 
\[
X_k = \frac{d}{da^k} \exp(a^i X_i)_{|_{a^i=0}} \;.
\]
A word of caution :
The Lie product is a commutator with respect to the
formal product $XY$, {\em i.e.} $[X,Y]=XY - YX$, which 
is however not an element of the $\mathfrak{g}$ or
$G$. It is straightly speaking not otherwise defined 
mathematically unless within the context of the 
corresponding universal enveloping algebra or 
a representation setting as a matrix/operator product.

A subgroup  is a part of the group which makes a group 
in itself.  A Lie subgroup $H$ of a Lie group is associated
to a Lie subalgebra  $\mathfrak{h}$ of  $\mathfrak{g}$.
In general, a subgroup  can be used to divide the group 
into a, possibly infinite, number of distinct cosets. For 
the case of a Lie group $G$, which can in itself obviously 
be seen as geometric space, the collection of cosets of
a Lie subgroup $H$ also can be seen as a geometric space, 
the coset space denoted by $G/H$, with each coset taken 
as an abstract point. A good picture of that has been 
presented in the main text. Note that coset spaces are 
not necessarily flat, {\em i.e.} may not be vector spaces,
but always homogeneous, {\em i.e.} look the same 
from every single point inside.

\end{appendices}

\acknowledgments
The authors wish to thank the students from their 
\textit{Group Theory and Symmetry} course at the
National Central University, in which 
a preliminary version of the above material was first 
prepared as a supplementary lecture note.  The
work is partially supported by research grant number
109-2119-M-008-016 of the MOST of Taiwan.

\end{document}